\documentclass[sn-mathphys]{sn-jnl}% Math and Physical Sciences Reference Style
\usepackage{bm}
\usepackage{color}
\usepackage{amsfonts}
\usepackage{ulem}
\usepackage{subfigure}

\def\br{\bm{\rho}}

\def\br{\bm{\rho}}

\jyear{2021}%

\theoremstyle{thmstyleone}%
\theoremstyle{thmstyletwo}%

\theoremstyle{thmstylethree}%

\begin{document}

\title[Comparative analysis of SNR in CPI architectures]{Comparative analysis of signal-to-noise ratio in correlation plenoptic imaging architectures}


\author[1,2]{\fnm{Gianlorenzo} \sur{Massaro}}
\equalcont{These authors contributed equally to this work.}
\author[3,4]{\fnm{Giovanni} \sur{Scala}}
\equalcont{These authors contributed equally to this work.}
\author*[1,2]{\fnm{Milena} \sur{D'Angelo}}\email{milena.dangelo@uniba.it}
\author[1,2]{\fnm{Francesco V.} \sur{Pepe}}

\affil*[1]{\orgdiv{Dipartimento Interateneo di Fisica}, \orgname{Universit\`a degli Studi di Bari}, \orgaddress{\city{Bari}, \postcode{70125}, \country{Italy}}}

\affil[2]{\orgdiv{Istituto Nazionale di Fisica Nucleare}, \orgname{Sezione di Bari}, \orgaddress{\city{Bari}, \postcode{70125}, \country{Italy}}}

\affil[3]{\orgdiv{International Centre for Theory of Quantum Technologies}, \orgname{University of Gdansk}, \orgaddress{\city{Gdansk}, \postcode{80-308}, \country{Poland}}}

\affil[4]{\orgdiv{Faculty of Physics}, \orgname{University of Warsaw}, \orgaddress{\city{Warsaw}, \postcode{02-903}, \country{Poland}}}


\abstract{Correlation plenoptic imaging (CPI) is a scanning-free diffraction-limited 3D optical imaging technique exploiting the peculiar properties of correlated light sources. CPI has been further extended to samples of interest to microscopy, such as fluorescent or scattering objects, in a modified architecture named correlation light-field microscopy (CLM). Interestingly, experiments have shown that the noise performances of CLM are significantly improved over the original CPI scheme, leading to better images and faster acquisition. In this work, we provide a theoretical foundation to such advantage by investigating the properties of both the signal-to-noise and the signal-to-background ratios of CLM and the original CPI setup.}




\maketitle

\section{Introduction}\label{sec:introduction}

    Plenoptic imaging is based on retrieving both spatial distribution and propagation direction of light within the single exposure of a digital camera \cite{adelson,ng,lippmann}; this enables both refocusing, in post-processing, and scanning-free 3D imaging. The technique is currently employed in several applications that include microscopy \citep{microscopy1,microscopy2,microscopy3,microscopy4}, particle image velocimetry \citep{piv}, wavefront and remote sensing \citep{thesis_wu,eye,atmosphere1,atmosphere2,remotesensing}, particle tracking and sizing \citep{tracking}, and stereoscopy \citep{adelson,muenzel,levoy}.
    Being capable of acquiring multiple 2D viewpoints of the sample of interest within a single shot, plenoptic apparata are amongst the fastest devices for performing scanning-free 3D imaging \citep{3dimaging}, as already shown recently for surgical robotics \citep{surgery}, imaging of animal neuronal activity \citep{microscopy4}, blood-flow visualization \citep{piv2}, and endoscopy \citep{endoscopy}. 
    The novelty of correlation plenoptic imaging (CPI) \citep{cpi_prl,cpi_qmqm,cpi_jopt,cpi_exp,cpi_technologies}, compared to conventional plenoptic imaging devices, is to employ a correlated light source enabling to collect information on the intensity distribution and the propagation direction of light on two distinct sensors, rather than one \citep{ng,georgiev1,georgiev2}. The most immediate advantage is to overome the strong tradeoff between spatial and directional resolution affecting conventional plenoptic imaging implementations  \citep{pittman,qu_superres,qu_superres2,sofi,undetected,dangelo_kim,scarcelli_er,tamma,genovese_review}.
    CPI, however, has the main drawback of no longer being a single-shot technique. In fact, in order to retrieve the amount of information required to reconstruct the 3D scene of interest, an average of the correlations between intensity fluctuations separately measured by the two sensors must be performed over many frames; each frame representing a different statistical realization of the correlated light source, which is generally chisen to be a source of entangled beams or chaotic light. To aim at real-time imaging the number of required frames should thus be reduced as much as possible. On the other hand, the number of statistical realizations that are sampled cannot be reduced too much without negatively affecting the image quality, namely, its signal-to-noise ratio (SNR) and signal-to-background ratio (SBR). Understanding the relationship linking the image quality with the number of  acquired frames would thus provide a powerful tool for  optimizing the measurement time of CPI and determining the necessary number of frames for reaching the desired output quality. This has been done in Ref.~\cite{cpi_snr}, where the  noise properties of the original CPI scheme \citep{cpi_prl,cpi_exp} (therein referred to as SETUP1) have been outlined.

    In the present work, we offer a detailed study of the SNR and the SBR characterizing correlation light-field microscopy (CLM), a versatile CPI architecture oriented to microscopy \cite{cpm_theory}. As we shall better explain in the next section, CLM enables extending the use of CPI to both self-emitting (\textit{e.g.} fluorescent) and scattering samples, and has the potential to be more robust to the presence of turbulence in the surroundings of the sample. Additionally, experimental evidence has shown that CLM is characterized by far superior noise performances compared to the original CPI scheme \citep{cpm_exp}. In this paper, we develop a solid theoretical background for interpreting and understanding this effect by comparing the expected noise performances of CLM with the ones of the original CPI scheme \citep{cpi_prl,cpi_exp,cpi_snr}.The paper is organized as follows. In Section~\ref{sec:results}, we report the analysis of the SNR and SBR characterizing both the unprocessed point-to-point correlation function of CLM and its final refocused image; the results are compared with one obtained for the original CPI architecture \citep{cpi_snr}. In Section~\ref{sec:discussion}, we discuss the theoretical and practical implications of the presented results, while in Section~\ref{sec:methods}, we report their theoretical and methodological basis. For the reader's convenience, lengthy computations and exact formulas are reported in the Appendix.
    
%%%%%%%%%%%%%%%%%%%%%%%%%%%%%%%%%%%%%%%%%%

\section{Results}\label{sec:results}

\subsection{CLM: working principle, correlation function and refocusing algorithm}

\begin{figure}
		\subfigure[Setup $\mathcal{G}$]{\includegraphics[width=0.45\textwidth]{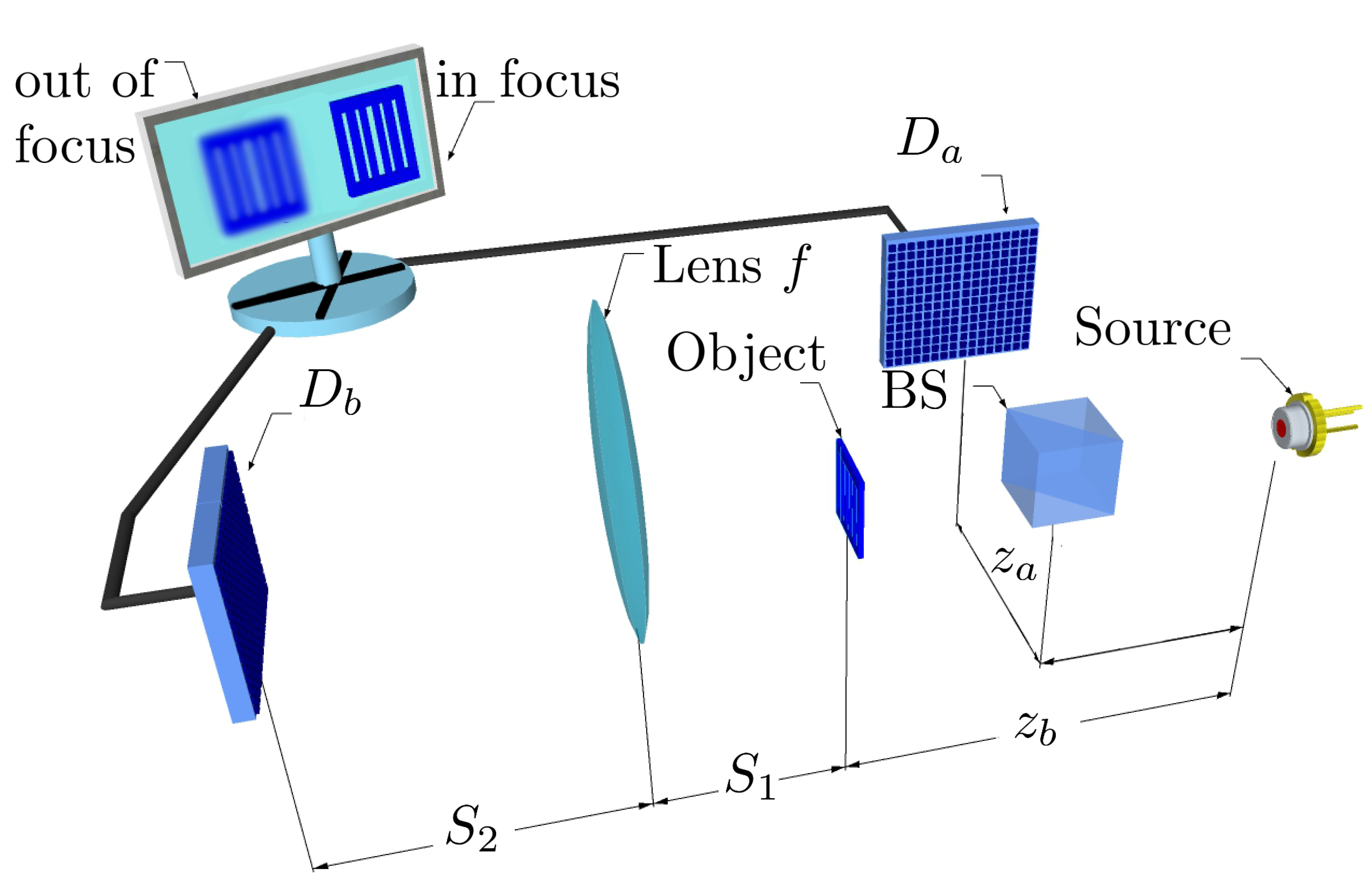}} \subfigure[Setup $\mathcal{M}$]{\includegraphics[width=0.45\textwidth]{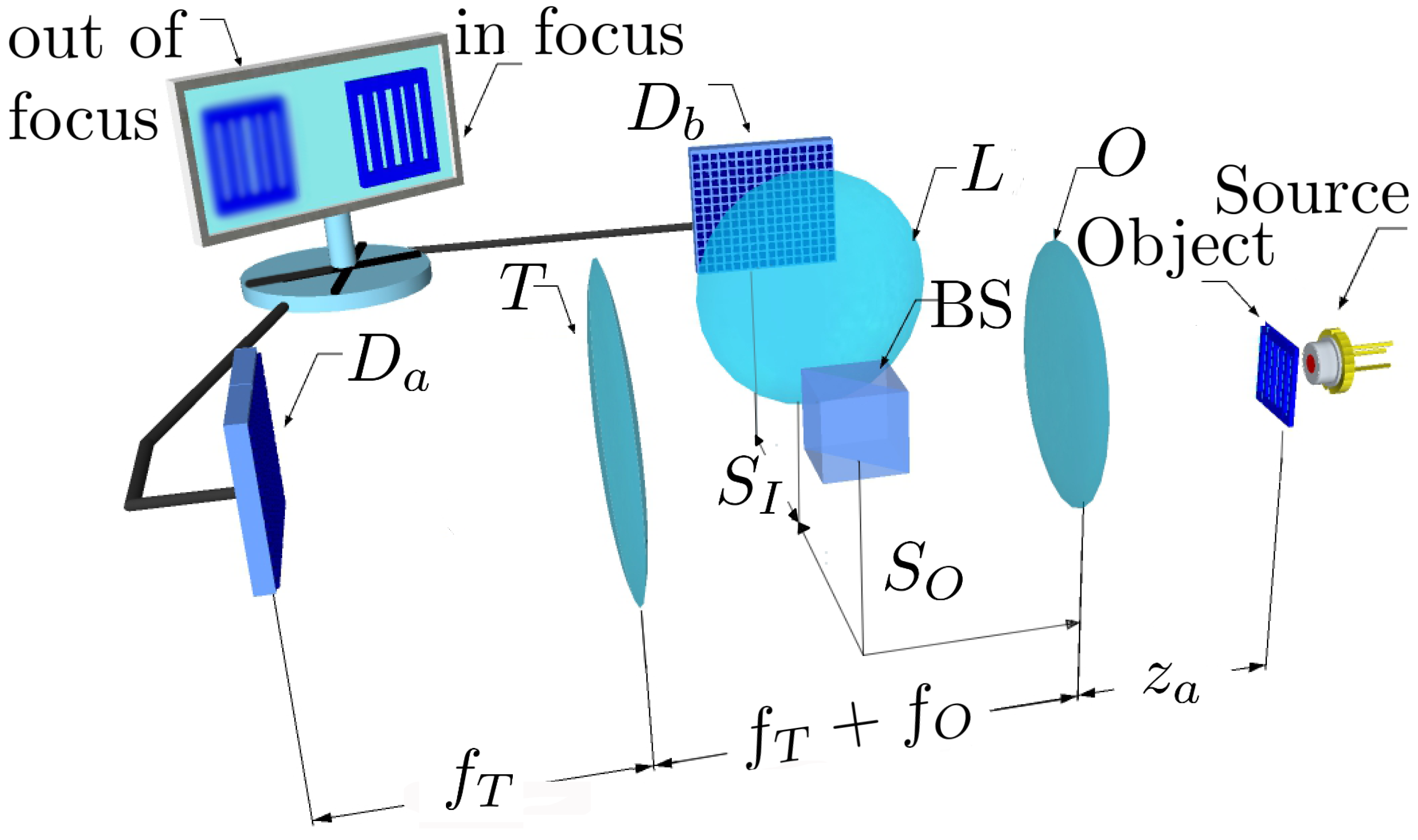}}
		\caption{Schematic representation of the original CPI, named $\mathcal{G}$ and based on so-called ghost imaging, and correlation light-field microscopy (CLM), named $\mathcal{M}$. See details in the text.
		}
		\label{fig:setups}
	\end{figure}

	Let us start by briefly presenting the two CPI architectures that will be compared throughout the paper: The original CPI scheme and CLM, as proposed in Ref.~\citep{cpi_prl} and \citep{cpm_theory}, respectively, and reported in Fig. ~\ref{fig:setups}(a) and (b). As emphasized in the figure, throughout the whole paper, the two architectures will be identified with the symbols $\mathcal{G}$ and $\mathcal{M}$, respectively.
	In both setups, light emitted by a chaotic source is split into two paths $a$ and $b$ by means of a beam-splitter (BS) and is recorded at the end of each path by the high-resolution detectors $D_{a}$ and $D_{b}$. Despite the intrinsic differences between the two schemes, the roles of $D_{a}$ and $D_{b}$ are similar: in both cases, $D_{b}$ is dedicated to retrieving information about the direction of propagation of light, while $D_{a}$ collects an image of the sample that can be either focused or out of focus, depending on where the latter is placed along the optical axis. The crucial difference lies in the fact that a conventional image of the object (i.e., based on intensity measurement) is available on $D_a$ in scheme $\mathcal{M}$, but cannot be retrieved in  scheme $\mathcal{G}$. In fact, the original CPI scheme is based on so-called ghost imaging \citep{GI_valencia}: the focused image of the object can only be retrieved by measuring correlations between a bucket (i.e., single-pixel) detector  (here represented by the entire sensor $D_{b}$, upon integration over all pixels) collecting light scattered/reflected/transmitted by the object, and a high-resolution detector (here $D_{a}$) placed at distance from the source equal to the object to source distance. However, as we shall see, in both setups, if the object is out of focus, the focused image can still be recovered by means of a refocusing algorithm applied to the point-by-point correlation of the intensity fluctuations measured by $D_{a}$ and $D_{b}$. 
	
	In scheme $\mathcal{G}$, light from a chaotic source is split by a beam splitter (BS). The transmitted beam illuminates the object, passes through an imaging lens, and is recorded by detector $D_{b}$, which is placed in the conjugate plane of the source, as generated by lens L. The reflected beam impinges on detector $D_{a}$. When the object to source distance is equal to the distance between $D_a$ and the source ($z_a=z_b$), its focused ghost imaging can be retrieved through correlations between each pixel of $D_a$ e the overall (integreted) signal from $D_b$. When the object to source distance differs from the $D_a$ to source distance ($z_a\neq z_b$), the ghost image is blurred, but can be refocused by reconstructing the point-to-point correlation between each pixel of $D_a$ and $D_b$ and applying a proper refocusing algorithm. Scheme $\mathcal{M}$, instead, belongs to a novel subclass of CPI schemes in which the object is placed before the beam-splitter, is modeled as a chaotic source, and can be transmissive, reflective, scattering, or even self-emitting. The setup involves an objective lens (O), a tube lens (T), and an auxiliary lens (L), with focal lengths $f_{O}$, $f_{T}$, $f_{L}$ respectively. Detector $D_{a}$ is fixed in the back focal plane of the tube lens, while the sample is at an arbitrary distance $z_{a}$ from the objective lens; its  conventional image is directly focused on $D_a$ when $z_{a}=f_{O}$, and is blurred otherwise. Detector $D_{b}$ retrieves the image of the objective lens, as produced by the lens L. Also in this setup, correlating intensity fluctuations on the two detectors provides the required plenoptic information enabling to reconstruct the focused image of the object also when the object is placed away from the conjugate plane of detector $D_a$, as defined by the objective and the tube lens.
	
 We shall now demonstrate that CLM entails an improvement of the noise properties with respect to the original CPI protocol. To this end, we shall consider the pixel-by-pixel correlation of the intensity fluctuations measured on the two disjoint detectors $D_{a}$ and $D_{b}$. More specifically, the intensity patterns $I_{A}\left(\boldsymbol{\rho}_{a}\right)$
	and $I_{B}\left(\boldsymbol{\rho}_{b}\right)$, with $\boldsymbol{\rho}_{a,b}$ the transverse coordinate on each detector plane, are recorded in time to reconstruct the correlation function
	\begin{equation}
		\Gamma (\boldsymbol{\rho}_{a},\boldsymbol{\rho}_{b})=\left\langle \Delta I_{A}(\boldsymbol{\rho}_{a})\Delta I_{B}(\boldsymbol{\rho}_{b})\right\rangle ,\label{Gamma1}
	\end{equation}
	where $\Delta I_{K}(\boldsymbol{\rho}_{k})=I_{K}(\boldsymbol{\rho}_{k})-\left\langle I_{K}(\boldsymbol{\rho}_{k})\right\rangle $ for $K=A,B$ and $k=a,b$ are the fluctuations of intensities, with respect to their average values, registered on each pixel of the two detectors.
	The expectation values in Eq.~\eqref{Gamma1} and in the average intensity should be evaluated considering the source statistics, but they can be approximated by time averages for stationary and ergodic	sources \citep{mandel}.

In both setups, the correlation function of Eq. \eqref{Gamma1} at a fixed $\boldsymbol{\rho}_{b}$ encodes a coherent image of the object \cite{cpi_prl,cpm_theory,coherent1, coherent2}. Each point $\boldsymbol{\rho}_{b}$ on the plane of detector $D_{b}$ corresponds to a different point of view on the scene (as further discussed in Sec. \ref{sec:Correlation-function}), and the images corresponding to different $\boldsymbol{\rho}_{b}$ coordinates are shifted one with respect to the other. If the position of the object in each setup satisfies a specific focusing condition ($z_a=z_b$ for $\mathcal{G}$, $z_a=f_O$ for $\mathcal{M}$), the relative shift of such images vanishes, and one can integrate over the detector $D_{b}$ to obtain an incoherent image characterized by an improved SNR with respect to the single images. However, if the (intensity or correlation-based) image of the object is out of focus, before piling-up the single coherent images through integration over $D_{b}$, they must be realigned according to the following refocusing algorithm:
\begin{equation}
	\Sigma_{\mathrm{ref}}(\boldsymbol{\rho}_{a})= \int \mathrm{d}^{2}\boldsymbol{\rho}_{b} \Gamma \left(\alpha\boldsymbol{\rho}_{a}+\beta\boldsymbol{\rho}_{b}, \boldsymbol{\rho}_{b}\right) , \label{Sigmaref}
\end{equation}
where the parameters $(\alpha,\beta)$ depend on the specific CPI architecture:
\begin{equation}
	(\alpha,\beta)=\left\{ \begin{matrix}{\displaystyle \left(\frac{z_{b}}{z_{a}}\,,\frac{1}{M}\left(\frac{z_{b}}{z_{a}}-1\right)\right)} & \text{for } \mathcal{G}  \\ \, \\ \displaystyle
		\left(-\frac{f_{O}}{z_{a}}M_a\,,\left(\frac{f_{O}}{z_{a}}-1\right)
		\frac{M_a}{M_b}\right) & \text{for } \mathcal{M},
	\end{matrix}\right.\label{refocus}
\end{equation}
where $M = S_2/(S_1+z_b)$ is the magnification of the image of the source on $D_b$, in setup $\mathcal G$, and $M_a=f_T / f_O$ and $M_b=S_I/S_O$ are the magnifications in arm $a$ and in arm $b$, respectively, in setup $\mathcal M$. It is easy to verify that in the focused cases ($z_a=z_b$ for $\mathcal{G}$, $z_a=f_O$ for $\mathcal{M}$), there is no need to either shift or scale the variables since the coherent images are already aligned and the high-SNR incoherent image can be obtained by simply integrating over $D_b$.

\subsection{Noise properties of the refocused images in the two CPI architectures}

Information on the noise affecting the refocused images can be obtained by evaluating the variance $\mathcal{F}(\br_a)$ of the physical quantity integrated in Eq.~\eqref{Sigmaref} (see Section~\ref{sec:methods} for derivation and properties). If a sequence of $N_{f}$ frames is collected in order to evaluate the refocused image, the root-mean-square error affecting the evaluation of this quantity can be estimated by $\sqrt{\mathcal{F}\left(\boldsymbol{\rho}_{a}\right)/N_{f}}$.
Therefore the signal-to-noise ratio reads
\begin{equation}
	\mathrm{SNR}_{\Sigma}\left(\boldsymbol{\rho}_{a}\right)=\frac{\Sigma_{\mathrm{ref}}\left(\boldsymbol{\rho}_{a}\right)}{\sqrt{\mathcal{F}\left(\boldsymbol{\rho}_{a}\right)/N_{f}}}.\label{Wick-1}
\end{equation}
However, due to its intrinsically local nature, the SNR cannot be used as an indicator of the quality of the refocused image in its entirety. We thus introduce a global figure of merit, the \textit{signal-to-background} ratio, defined as:
\begin{equation}\label{SBR_sigma}
	\text{SBR}_{\Sigma}=\frac{\Sigma_{\mathrm{\text{sig}}}}{\sqrt{ \Sigma_{\mathrm{\text{back}}}^{2} + \mathcal{F}_{\text{back}}/N_{f}}},
\end{equation}
where $\Sigma_{\mathrm{\text{sig}}}$ is the refocusing function, evaluated in a point where signal is expected, and the quantities in the denominators are reference values of $\Sigma$ and $\mathcal{F}$, evaluated for the \textit{background}, namely, those parts of the image in which essentially no signal is expected. 
The distinction between SNR and SBR becomes relevant whenever the variance of the refocused image depends on the signal magnitude, with the two being substantially identical when uniform noise throughout the image is expected (as, \textit{e.g.}, in ghost imaging).
The analytical expressions for SNR and SBR in the two CPI architectures of interest are quite cumbersome even in a geometrical-optics approximation, and enable only a few qualitative considerations. To overcome this difficulty and get a better insight about the differences in noise performances of the two schemes, numerical simulations have been carried over.

\subsubsection{Simulations of refocused images in the two CPI architectures}\label{subsec:Choice-of-parameters}

\begin{figure}
    \centering
    \includegraphics[width=0.45\textwidth]{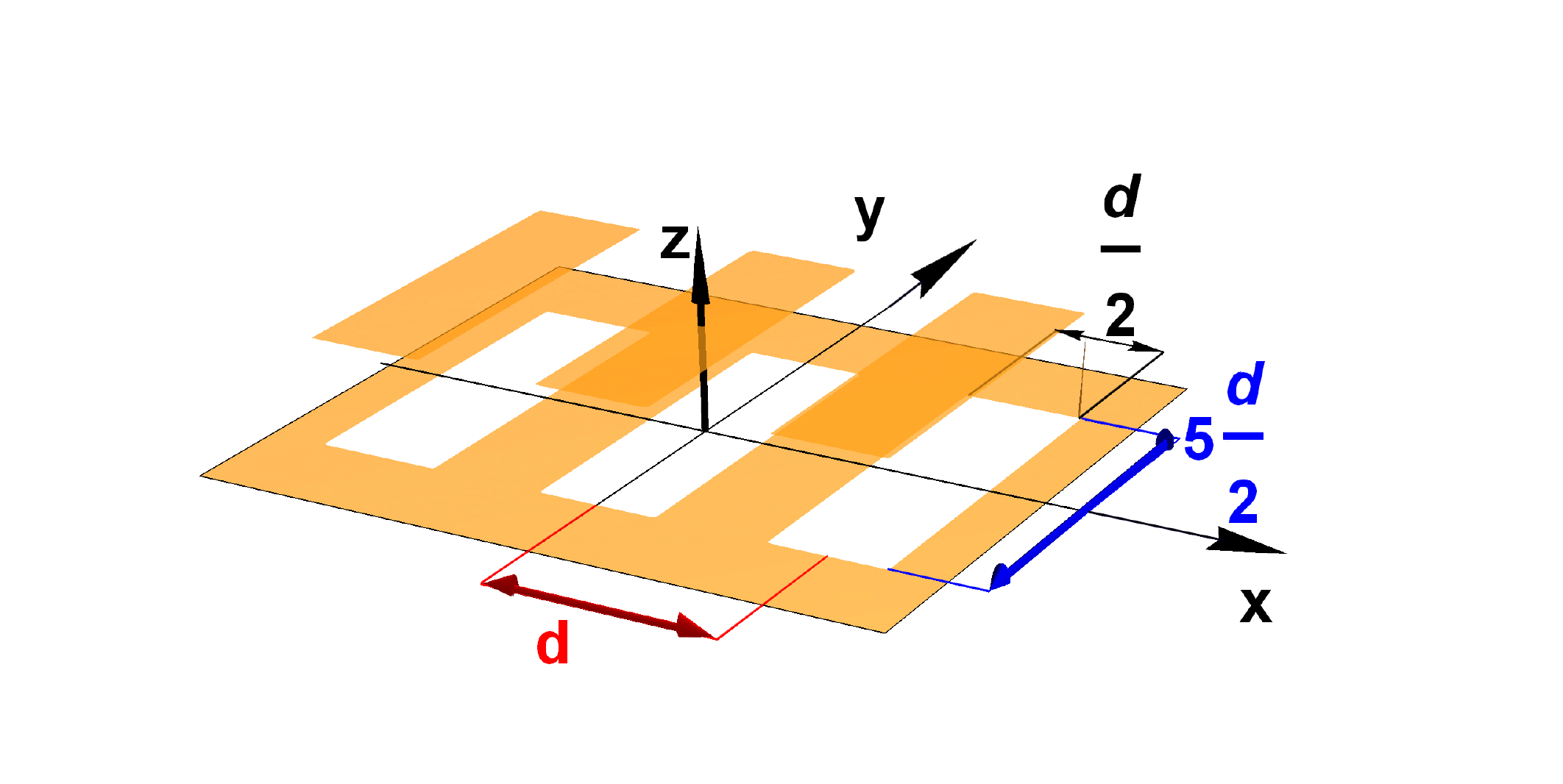}
	\hspace{.03\textwidth}
	\includegraphics[width=0.45\textwidth]{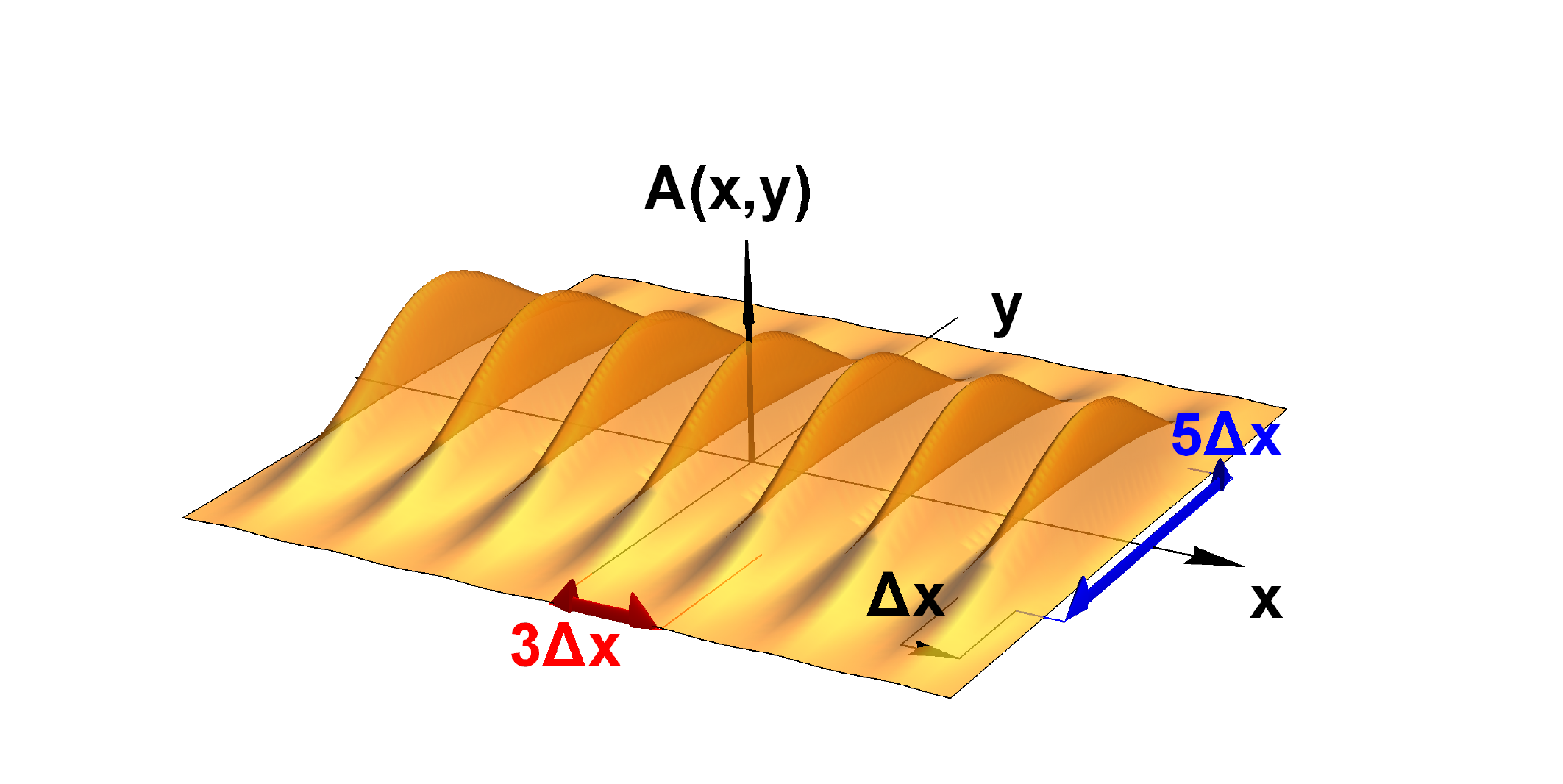}
    \caption{On the left hand side, the object binary transfer function used for the simulations to which Fig\ref{fig:snr_triple_slit-1} and \ref{fig:snr_triple_slit} refer: three equally spaced transimissive slits with varying parameter $d$; on the right hand side, the gaussian object transfer function of Eq. \eqref{eq:object-1}, used for simulations in which the number of details is varied from 1 to 8.}
    \label{fig:obj}
\end{figure}

Here we shall discuss the results of the numerical simulations of the noise performances obtained in the two CPI architectures under consideration, for two different objects (see Fig. \ref{fig:obj}): a set of three equally spaced and uniformly transmissive slits, analogous to a typical element of a negative resolution test target, and a set of $N$ equally spaced Gaussian slits, with $N$ varying from 1 to 8. Specifically, the latter masks are characterized by field transmittance profile
\begin{equation}\label{eq:object-1}
	A\left(x,y\right)=\left[\sum_{i=1}^N f\left(x;x_{i},w_{x}\right)\right]\cdot f\left(y;0,w_{y}=5w_{x}\right),
\end{equation}
with 
\begin{equation}
    f\left(x;x_{i},w\right)=\exp\left(-\frac{\left(x-x_{i}\right)^{2}}{2w^{2}}\right)
\end{equation}
and $\{x_i\}_{1\leq i\leq N}$ the set of slit centers along $x$. 

The noise performances of the two architectures will be compared as the transmissive area of the two objects is changed, which is obtained by varying the slit size for the first object, and $N$ for the second. In the latter case, variation of $N$ also enables checking the capability of the two setups to resolve an increasing number of details. In fact, the noise properties of ghost imaging depend only on the total transmissive area of the object, and not on the number of details \cite{GI_valencia}. Considering the difference in the image formation in CLM and in the (ghost imaging-based) original CPI scheme, it is interesting to investigate this property also in CPI architectures under investigation.

\begin{figure}
\centering
	\includegraphics[width=0.5\textwidth]{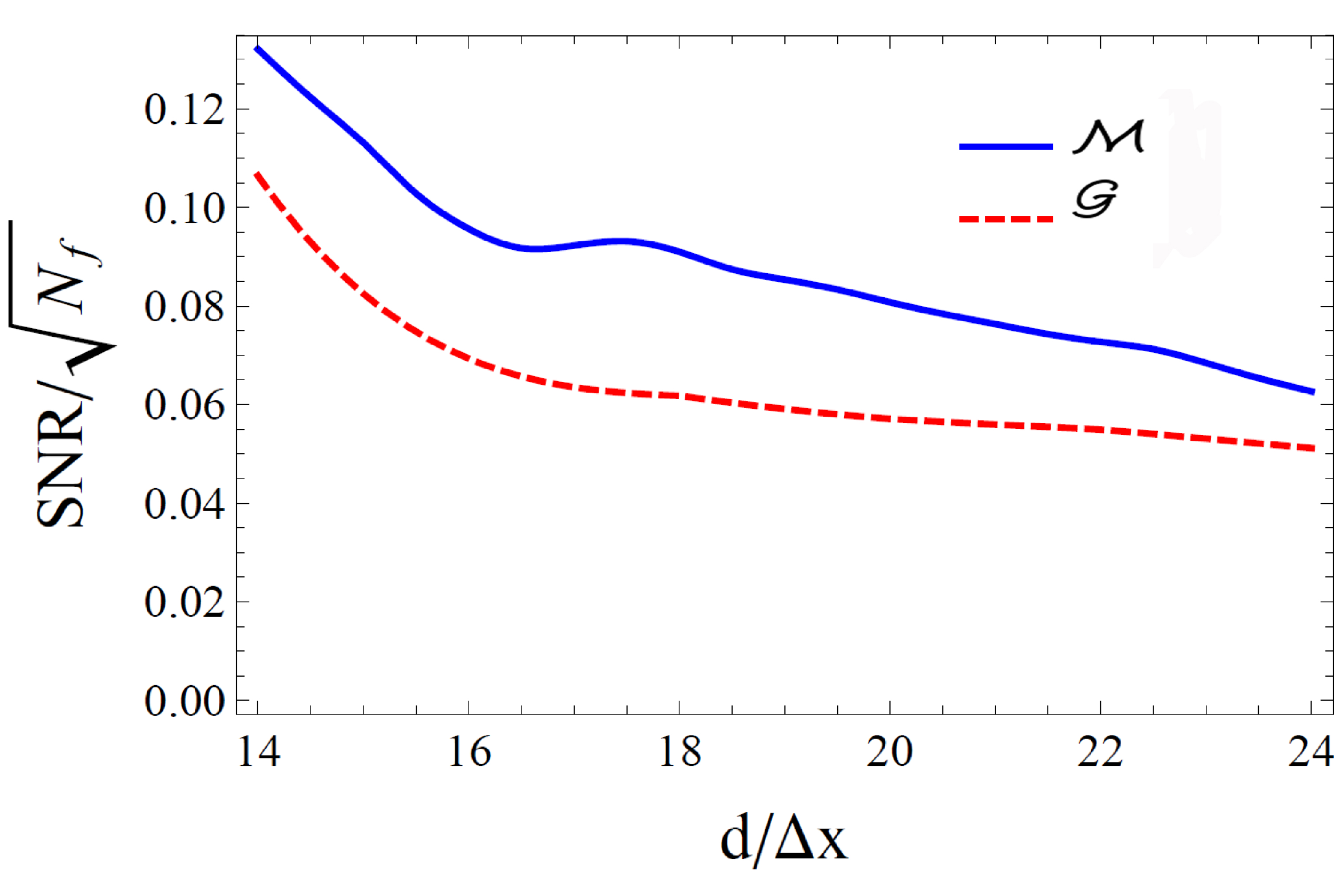}
	\caption{SNR of the refocused images $\Sigma_{\mathrm{ref}}$ of the triple slit reported in Fig. \ref{fig:obj}, placed $1\,$mm away from the plane at focus, averaged over the central points of the three slits and normalized with the square root of the frame number $N_f$, evaluated for the $\mathcal{M}$ (solid blue line) and $\mathcal{G}$ (dashed red line) architectures characterized by the same numerical aperture, hence, resolution of the focused image.
	}
	\label{fig:snr_triple_slit-1}
\end{figure}

Let us start by considering the refocused images of a planar mask made of three parallel transmissive slits, as from Fig. \ref{fig:obj}(a). In order to match the parameters for $\mathcal{G}$ and $\mathcal{M}$, we consider in both cases a mask placed at a distance of $26.6\,\mathrm{mm}$ from the focusing element, namely $1\,\mathrm{mm}$ farther than the conjugate plane of the sensor.
The resolution at focus for $\mathcal{M}$ is defined
by the typical point-spread function of the microscope \cite{cpm_theory}.  On the other hand, $\mathcal{G}$ is not capable of imaging the object with intensity
measurements, but only through ghost imaging, namely by correlating
the total intensity collected by $D_{b}$ (regarded as a bucket detector)
and the signals acquired by each pixel of $D_{a}$. Therefore, as in a focused ghost imaging system, the output image is given by
\begin{equation}\label{eq:ghostres-1}
	\Sigma(\bm{\rho}_{a})= 
	\int d^{2}\rho_{b}\ 
	\Gamma(\bm{\rho}_{a},\bm{\rho}_{b}) = 
	\int d^{2}\rho^\prime\ 
	\lvert A(\bm{\rho}^\prime)\rvert^{2}
	\text{PSF}_{\text{CPI}}
	(\bm{\rho}^{\prime}-\bm{\rho}_{a}) ,
\end{equation}
so that the resolution at focus can be defined through the positive point-spread
function $\mathrm{PSF_{CPI}}$ (see \cite{cpi_prl}) as in conventional imaging. The resolution can be made equal for the two setups by matching their numerical apertures and choosing the source-to-$D_a$ distance in $\mathcal{G}$ to be equal to the objective focal length in $\mathcal{M}$.
Moreover, the two objects are illuminated by chaotic light of the same wavelength $\lambda=550\,\mathrm{nm}$, while the numerical apertures of the imaging devices are equal and gives a resolution $\Delta x = 3\,\mu\mathrm{m}$ on the plane in which the correlation image is focused. In the comparison reported in Figure~\ref{fig:snr_triple_slit-1}, we can observe that the performance of setup $\mathcal{M}$ is better for all the considered values of the center-to-center slit distance, providing evidence of the advantage entailed by this setup.

Let us now consider the noise performance of $\mathcal{G}$ and $\mathcal{M}$ in the case of the second object, while varying the number of slits $N$, and considering both focused and refocused images. In both protocols, focused images are characterized by diffraction- limited resolution, which we set to $\Delta x=1.0\,\mu\mathrm{m}$ in both setups. In this condition, the term $\mathcal{F}_{0}$ in $\mathcal{M}$ has the strongest
spatial modulation, while it smooths out as the object is moved away
from the focused plane. Such a spatial dependence, which is not present
in the corresponding term for $\mathcal{G}$, is expected to have great impact
on the SBR.

\begin{figure}
\centering
	\includegraphics[width=0.45\textwidth]{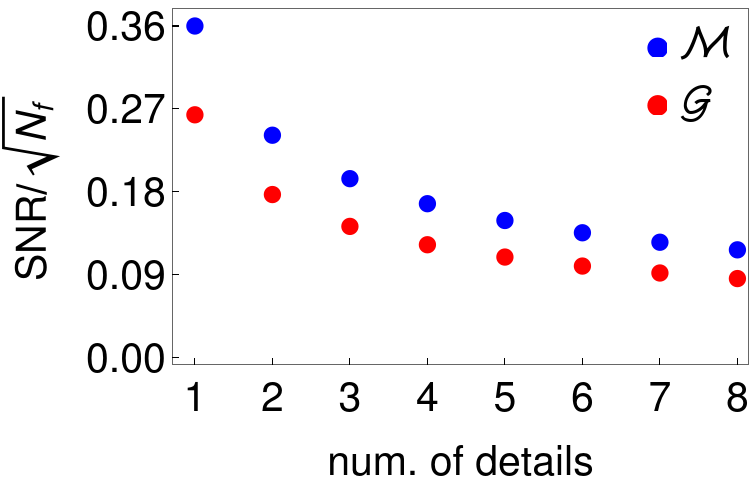}
	\hspace{.03\textwidth}
	\includegraphics[width=0.45\textwidth]{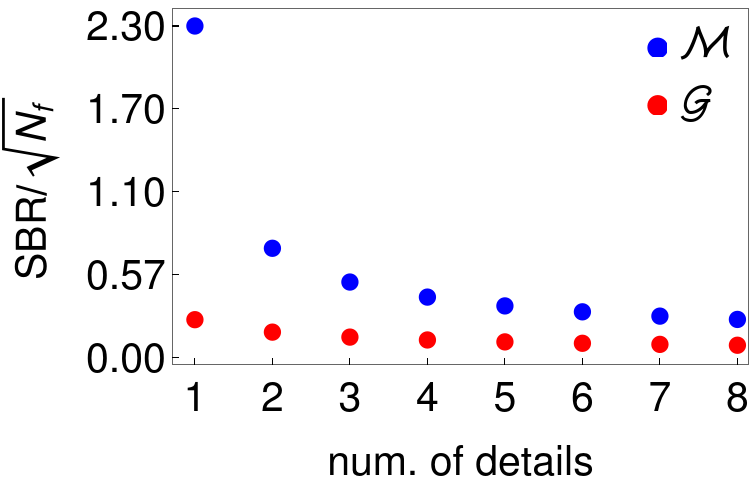}
	\caption{Comparison of the SNR (left panel) and SBR (right panel) of the final image obtained from Eq. \ref{eq:Sigma_ref}. The plots refer to the case of focused images, obtained in $\mathcal{G}$ (red dots) and $\mathcal{M}$ (blue dots) schemes, for a varying number of details within an object with transmittance profile reported in Eq. \eqref{eq:object-1} and Fig. \ref{fig:obj} (b).}
	\label{fig:sigma-num_slit-1}
\end{figure}

\begin{figure}
\centering
	\includegraphics[width=0.5\textwidth]{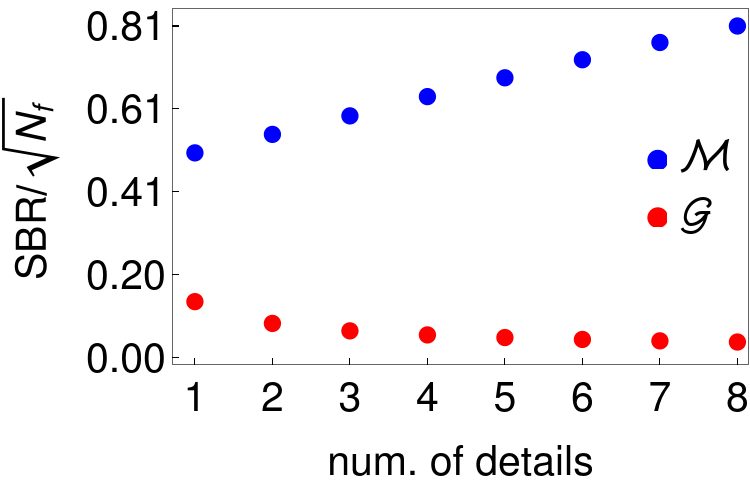}
	\caption{Comparison of the SBR of the final image obtained from Eq. \ref{eq:Sigma_ref} in the case of a deeply out-of-focus object. The images are obtained in setups $\mathcal{G}$ (red dots) and $\mathcal{M}$ (blue dots) for a varying number of Gaussian slits within an object with transmittance profile reported in Eq. \eqref{eq:object-1} and Fig. \ref{fig:obj} (b).}
	\label{fig:sigma-num_slit}
\end{figure}

In the case of focused images, the simulations reported in Figure~\ref{fig:sigma-num_slit-1} have been performed with the center-to-center slit distance $|x_{i}-x_{i-1}|=3\Delta x$ and width $w_x=\Delta x$, thus making the object details very close to the best achievable resolution in both setups. The SNR is averaged on the maxima of the image; to compute SBR, the signal and the background are evaluated by averaging, respectively, over the maxima and the minima of the image.

In the out-of-focus case, another feature of the refocusing functions has to be taken into account in order to rule out contributions that are not strictly related with the pure noise performance: a noise comparison can only make sense as long as the image quality is the same from an optical point of view, namely, if the details in the images are refocused with the same visibility
\begin{equation}
	\mathcal{V}=\frac{d^{+}-d^{-}}{d^{+}+d^{-}},
\end{equation}
where $d^{+}$ is the average value of the image on the peaks (corresponding to the slit centers) and $d^{-}$ the value in the
minima between adjacent peaks. In both setups, as the object
is moved away from the plane at focus, the visibility of refocused images decreases, and they do so with a slightly different trend in the two setups. Therefore, the slit distance in the out-of-focus regime
have been optimized so that, if the object is placed at the same distance
from the focused plane in the two setups, the refocusing visibility is 99\% in both setups. We also made sure to consider a case in which the object is very far from the plane at focus, choosing a distance from the focused plane equal to 30 times the natural depth of focus of the systems, which is the same for both setups. In this deeply out-of-focus regime, the SNR is calculated in the maxima of the refocused function and the SBR comes out to be identical to the SNR; we thus report in Figure~\ref{fig:sigma-num_slit} only the result for the SBR. In fact, while the noise profile in $\mathcal{G}$ is always constant, independently of the object distance, it is not constant in $\mathcal{M}$, where it flattens out the more the object is moved away from the conjugate plane of the detector, making no difference as to where the term $\mathcal{F}_{\text{back}}$ in Eq.~\eqref{SBR_sigma} is evaluated. On the other hand, the term $\Sigma_{\text{back}}$ tends to vanish in a resolved image. 

\begin{figure}
\centering
	\includegraphics[width=0.45\textwidth]{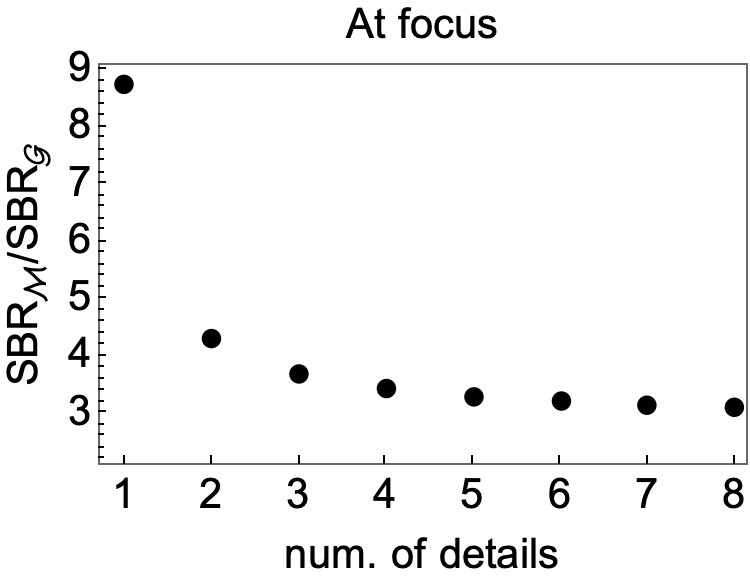}
	\hspace{.03\textwidth}
	\includegraphics[width=0.45\textwidth]{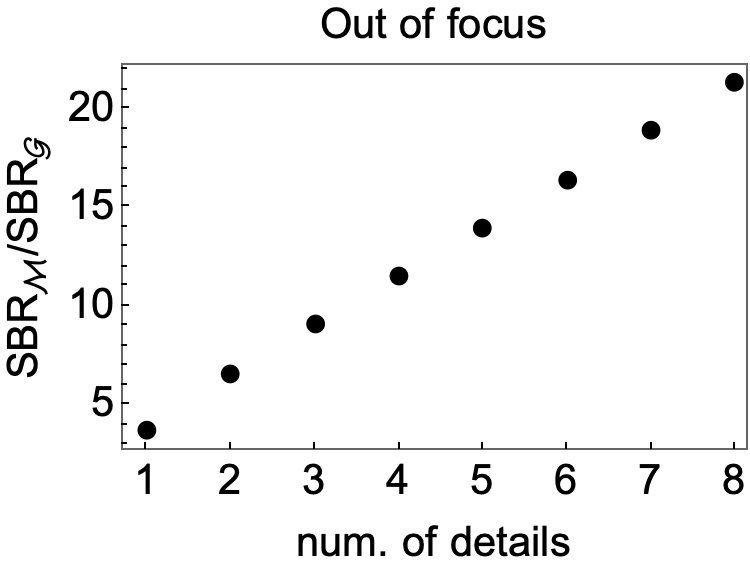}
	\caption{Ratio between the SBR of the final images obtained from Eq. \ref{eq:Sigma_ref} in setup $\mathcal{M}$ and $\mathcal{G}$, by varying the number of details of an object with transmittance profile given by Eq. \eqref{eq:object-1}. Left panel refers to the case at focus, while the right panel to the out-of-focus case.}
	\label{fig:sigma-num_slit-2}
\end{figure}

For a better quantification of the advantage of $\mathcal{M}$ over $\mathcal{G}$ in terms of SBR (which, in the out-of-focus case, also tends to coincide with SNR), we report in Figure~\ref{fig:sigma-num_slit-2} the ratio of the values obtained in the two setups.

\subsection{Noise properties of the correlation function in the two CPI architectures}\label{sec:Correlation-function}

The refocused image is a useful tool in any plenoptic imaging device, since it gathers all the information available from the collected data on the objects on a given plane, increasing the SNR and suppressing contributions from other planes. However, each 2D slice of $\Gamma(\br_a,\br_b)$ contains information on the object, providing, for each fixed $\br_b$, a high-depth-of-field image observed from a given point of view. If the object is planar and characterized by the field transmittance $A(\bm{\rho})$, the correlation function in a geometrical approximation
reads, up to irrelevant factors,
\begin{equation}\label{eq:multipers}
	\Gamma \left(\bm{\rho}_{a},\bm{\rho}_{b}\right) \simeq \lvert A(\alpha\bm{\rho}_{a}+\beta\bm{\rho}_{b})\rvert^{2n},
\end{equation}
where $\alpha$ and $\beta$ are reported in Eq.~\eqref{refocus}, $n=1$ for $\mathcal{G}$, and $n=2$ for $\mathcal{M}$. As anticipated, we see from
Eq.~\eqref{eq:multipers} that the correlation function, as
the detector points are varied, provides a shifted and scaled representation
of the aperture function of the sample. In particular, varying the
coordinate $\bm{\rho}_{b}$ is equivalent to varying the point of view
on the object imaged on the detector $D_{a}$. An intuitive explanation
of this effect is that the correlation with the pixels of $D_{b}$ automatically selects rays emitted by a well-defined part of the source, in $\mathcal{G}$, or passing through a well-defined part of the objective lens, in $\mathcal{M}$. When dealing with a 2D object, as in our case, refocusing might look like a useless operation, since the object profile is already contained in each single viewpoint. However, the need to refocus arises from the fact that single perspectives are typically too noisy for the object to be distinguished from the background: thus, summing them together improves the SNR and the SBR by the squared root of the number of statistically independent perspectives. Different from the first CPI experiment \cite{cpi_exp}, which was based on scheme $G$, in a recent CLM experiment \cite{cpm_exp}, single point-of-view images with high enough SNR for the objects to be recognized were obtained. Extending the study and comparison of noise performances directly to the correlation function rather than to the final refocused images is thus interesting and worth investigating.

We define the noise affecting the correlation function as the variance of
the product of intensity fluctuations on $D_{a}$ and $D_{b}$, divided by $\sqrt{N_f}$. In this case, there is no suppression of lower-order terms, so that all eight-point correlators contribute to the variance, which reads
\begin{align}
	\mathcal{F}_{\Gamma}(\boldsymbol{\rho}_{a},\boldsymbol{\rho}_{b})= & \left\langle \left(\Delta I_{A}\left(\boldsymbol{\rho}_{a}\right)\Delta I_{B}\left(\boldsymbol{\rho}_{b}\right)\right)^{2}\right\rangle -\Gamma\left(\boldsymbol{\rho}_{a},\boldsymbol{\rho}_{b}\right)^{2}\nonumber \\
		= & \left[\mathcal{I}_{A}(\boldsymbol{\rho}_{a})\mathcal{I}_{B}(\boldsymbol{\rho}_{b})+2\Gamma\left(\boldsymbol{\rho}_{a},\boldsymbol{\rho}_{b}\right)\right]^{2}-\Gamma\left(\boldsymbol{\rho}_{a},\boldsymbol{\rho}_{b}\right)^{2}. \label{eq:FGamma}
\end{align}
The SNR is thus defined as
\begin{equation}
	\mathrm{SNR}_{\Gamma}(\boldsymbol{\rho}_{a},\boldsymbol{\rho}_{b})=  \frac{\Gamma(\boldsymbol{\rho}_{a},\boldsymbol{\rho}_{b})}{\sqrt{\mathcal{F}_{\Gamma}(\boldsymbol{\rho}_{a},\boldsymbol{\rho}_{b})/N_{f}}}
	= \sqrt{N_{f}}\frac{1}{\sqrt{\left[\left(\frac{\mathcal{I}_{A}(\boldsymbol{\rho}_{a})\mathcal{I}_{B}(\boldsymbol{\rho}_{b})}{\Gamma(\boldsymbol{\rho}_{a},\boldsymbol{\rho}_{b})}+2\right)-1\right]}}.
\end{equation}
The definition of SBR can be generalized to the study of the correlation function
by considering the signal evaluated at a generic pair of points  $(\boldsymbol{\rho}_{a},\boldsymbol{\rho}_{b})$ and the background evaluated at a reference point $(\boldsymbol{\rho}_{a}^{\prime},\boldsymbol{\rho}_{b})$ in which the expected intensity of the considered point-of-view image is practically
vanishing:
\begin{equation}
	\mathrm{SBR}_{\Gamma}(\boldsymbol{\rho}_{a},\boldsymbol{\rho}_{b};\boldsymbol{\rho}_{a}^{\prime})= \Bigg(\frac{\Gamma(\boldsymbol{\rho}_{a},\boldsymbol{\rho}_{b})}{\sqrt{\mathcal{F}(\boldsymbol{\rho}_{a}^{\prime},\boldsymbol{\rho}_{b})/N_{f}}}\Bigg) _{\Gamma(\boldsymbol{\rho}_{a}^{\prime},\boldsymbol{\rho}_{b})\simeq0} = \sqrt{N_{f}}\frac{\Gamma(\boldsymbol{\rho}_{a},\boldsymbol{\rho}_{b})}{\mathcal{I}_{A}(\boldsymbol{\rho}_{a}^{\prime})\mathcal{I}_{B}(\boldsymbol{\rho}_{b})}.
\end{equation}
As evident, the SNR is determined by the adimensional quantity $\Gamma/\mathcal{I}_{A}\mathcal{I}_{B}$. In order to estimate this ratio, we consider the cases of focused
and out-of-focus image in the geometrical approximation, which is
of much simpler interpretation compared to the refocusing function,
as we shall see shortly below.

\subsubsection{Focused case}
	
For $\mathcal{G}$, the correlation $\Gamma(\boldsymbol{\rho}_{a},\boldsymbol{\rho}_{b})$
encodes the image of the object intensity transmission function $|A|^2$ with unitary magnification, while the intensity
on the detector $D_{a}$ is uniform. Taking, for definiteness, $\boldsymbol{\rho}_{b}=0$, corresponding to the central point of view, we obtain
\begin{equation}
	\Gamma(\boldsymbol{\rho}_{a},0)\sim\lvert A(\boldsymbol{\rho}_{a})\rvert^{2},\qquad\mathcal{I}_{A}(\boldsymbol{\rho}_{a})\sim\mathrm{const.},
\end{equation}
yielding 
\begin{equation}
	\mathrm{SNR}_{\Gamma}(\boldsymbol{\rho}_{a},0)\simeq\frac{\sqrt{N_{f}}}{\sqrt{\left(\frac{p_{\mathcal{G}}}{\lvert A(\boldsymbol{\rho}_{a})\rvert^{2}}+2\right)^{2}-1}},
	\qquad
	\mathrm{SBR}_{\Gamma}(\boldsymbol{\rho}_{a},0;\boldsymbol{\rho}_{a}^{\prime})\simeq\frac{\lvert A(\boldsymbol{\rho}_{a})\rvert^{2}}{p_{\mathcal{G}}},
\end{equation}
with $p_{\mathcal{G}}$ a $\br_a$-independent constant, related to the setup parameters and the object total transmittance. Therefore, both quantities vanish if the signal is small. 

For $\mathcal{M}$, both $\Gamma$ and the intensity on $D_{a}$ encode an image of the \textit{squared} object intensity profile, 
\begin{equation}
	\Gamma(\boldsymbol{\rho}_{a},0)\sim \bigg\lvert A\left(-\frac{\boldsymbol{\rho}_{a}}{M_a}\right)\bigg\rvert^{4},\qquad\mathcal{I}_{A}(\boldsymbol{\rho}_{a})\sim \bigg\lvert A\left(-\frac{\boldsymbol{\rho}_{a}}{M_a}\right)\bigg\rvert^{2},
\end{equation}
where $f_{O}$ is the objective focal length and $f_{T}$ the tube
lens focal length. Therefore, the SNR is approximately independent
of the spatial modulation of the signal 
\begin{equation}
	\mathrm{SNR}_{\Gamma}(\boldsymbol{\rho}_{a},0)\simeq\frac{\sqrt{N_{f}}}{\sqrt{\left(\frac{p_{\mathcal{M}}}{\lvert A(-\br_a/M_a)\rvert^2}+2\right)^{2}-1}};
\end{equation}
with $p_{\mathcal{M}}$ a $\br_a$-independent constant, related with the setup parameters and the object total transmittance.
On the other hand, since the intensity $\mathcal{I}_{A}$ vanishes
out of the object profile, we get
\begin{equation}
	\mathrm{SBR}_{\Gamma}(\boldsymbol{\rho}_{a},0;\boldsymbol{\rho}_{a}^{\prime})\to\infty ,
\end{equation}
provided $\lvert A(-(f_O/f_T)\br_a)\rvert$ is not vanishing.

\subsubsection{Out-of-focus case}

In the out-of-focus case, intensity on the spatial sensor remains
uniform in the case of $\mathcal{G}$,
\begin{equation}
	\Gamma(\boldsymbol{\rho}_{a},0)\sim\bigg\lvert A\left(\frac{z_{b}}{z_{a}}\boldsymbol{\rho}_{a}\right)\bigg\rvert^{2},\qquad\mathcal{I}_{A}(\boldsymbol{\rho}_{a})\sim\mathrm{const.}
\end{equation}
with $z_{a}$ the source-to-$D_{a}$ distance and $z_{b}$
the source-to-object distance, but tends to become uniform even in
the case of $\mathcal{M}$, where
\begin{equation}
	\Gamma(\boldsymbol{\rho}_{a},0)\sim \bigg\lvert A\left(-\frac{z_{a}}{f_{T}}\boldsymbol{\rho}_{a}\right)\bigg\rvert^{4}, \qquad\mathcal{I}_{A}(\boldsymbol{\rho}_{a})\sim\mathrm{const.}
\end{equation}
Therefore, one obtains formally similar results in the case of $\mathcal{G}$
\begin{align}
	\mathrm{SNR}_{\Gamma}(\boldsymbol{\rho}_{a},0) \simeq & \frac{\sqrt{N_{f}}}{\sqrt{\left(\frac{q_{\mathcal{G}}}{\lvert A(z_{b}\boldsymbol{\rho}_{a}/z_{a})\rvert^{2}}+2\right)^{2}-1}}, \\
	\mathrm{SBR}_{\Gamma}(\boldsymbol{\rho}_{a},0;\boldsymbol{\rho}_{a}^{\prime}) \simeq & \sqrt{N_{f}}\frac{\lvert A(z_{b}\boldsymbol{\rho}_{a}/z_{a})\rvert^{2}}{q_{\mathcal{G}}},
\end{align}
and $\mathcal{M}$ 
\begin{align}
	\mathrm{SNR}_{\Gamma}(\boldsymbol{\rho}_{a},0)  \simeq & \frac{\sqrt{N_{f}}}{\sqrt{\left(\frac{q_{\mathcal{M}}}{\lvert A(-z_{a}\boldsymbol{\rho}_{a}/f_{T})\rvert^{4}}+2\right)^{2}-1}}, \\
	\mathrm{SBR}_{\Gamma}(\boldsymbol{\rho}_{a},0;\boldsymbol{\rho}_{a}^{\prime})  \simeq & \sqrt{N_{f}}\frac{\lvert A(-z_{a}\boldsymbol{\rho}_{a}/f_{T})\rvert^{4}}{q_{\mathcal{M}}}.
\end{align}
In the above equations, $q_{\mathcal{G}}$ and $q_{\mathcal{M}}$ are constant with respect to the object transverse coordinates, depending instead on the setup parameters, on the object axial position and on the total transmittance.
Therefore, in the out of focus case, the possible advantage of $\mathcal{M}$ depends on the detailed structure of the intensities and the correlation function.

\subsubsection{Simulations of correlation functions}
	
We repeated the analysis of SNR and SBR for the correlation functions in the same modalities discussed in Section~\ref{subsec:Choice-of-parameters}. In all the cases that follow, the central viewpoint is analyzed, namely the correlation image formed by rays passing through the center of the focusing element, as encoded in $\Gamma(\br_a,\br_b=0)$. 

\begin{figure}
	\centering \includegraphics[width=0.5\textwidth]{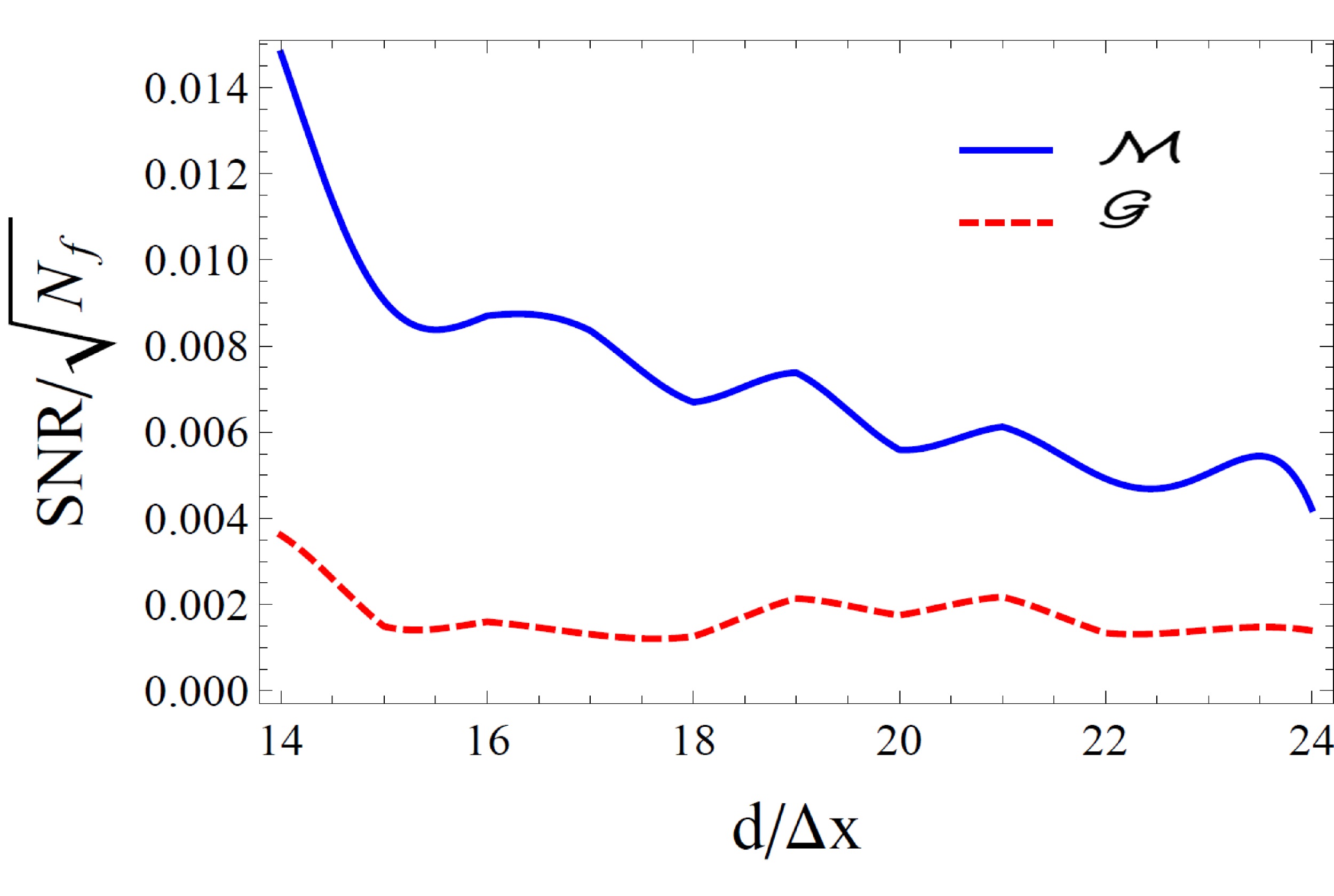}
	\caption{SNR of the central point-of-view images of the correlation function $\Gamma(\br_a,0)$ of the triple slit reported in Fig. \ref{fig:obj}, mediated over the central points of the three slits and normalized with the square root of the frame number $N_f$. The setup and object parameters are the same as those reported in Figure~\ref{fig:snr_triple_slit-1}. Solid blue line refers to setup $\mathcal{M}$, and dashed red line to $\mathcal{G}$. }
	\label{fig:snr_triple_slit}
\end{figure}

The study of SNR on the images of out-of-focus triple-slit masks is reported in Figure~\ref{fig:snr_triple_slit}, with the same parameter choice used to characterize noise in the refocused images. By comparing the results with those reported in Figure~\ref{fig:snr_triple_slit-1}, it is evident that, as expected, the SNR of the single viewpoint is lower than the SNR on refocused images. On the other hand, the relative advantage entailed by $\mathcal{M}$ over $\mathcal{G}$ is even amplified. 

\begin{figure}
\centering
	\includegraphics[width=0.45\textwidth]{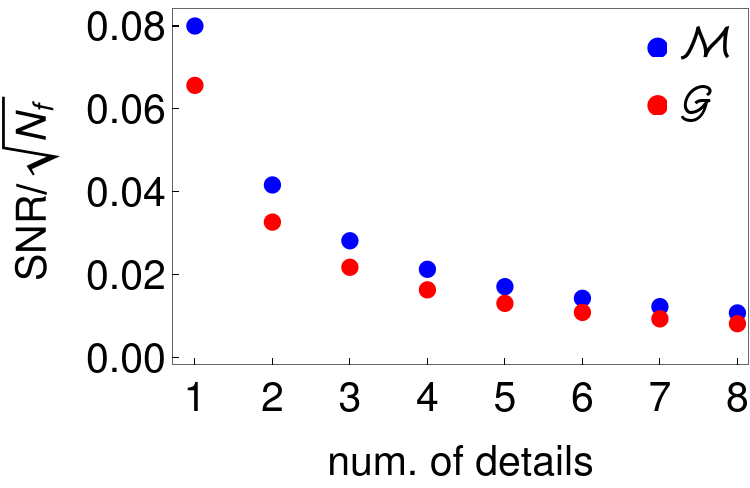}
	\hspace{.03\textwidth}
	\includegraphics[width=0.45\textwidth]{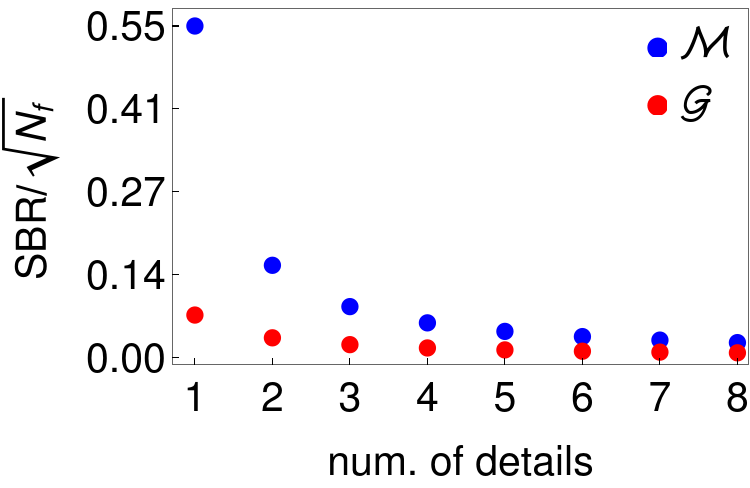}
	\caption{Comparison of the SNR (upper panel) and SBR (lower panel) at focus for the correlation function $\Gamma(\br_a,0)$ obtained in setups $\mathcal{G}$ (red dots) and $\mathcal{M}$ (blue dots), by varying the number of details (Gaussian slits) of an object with transmittance profile \eqref{eq:object-1}.
	}
	\label{fig:corr-num_slit}
\end{figure}

\begin{figure}
\centering
	\includegraphics[width=0.44\textwidth]{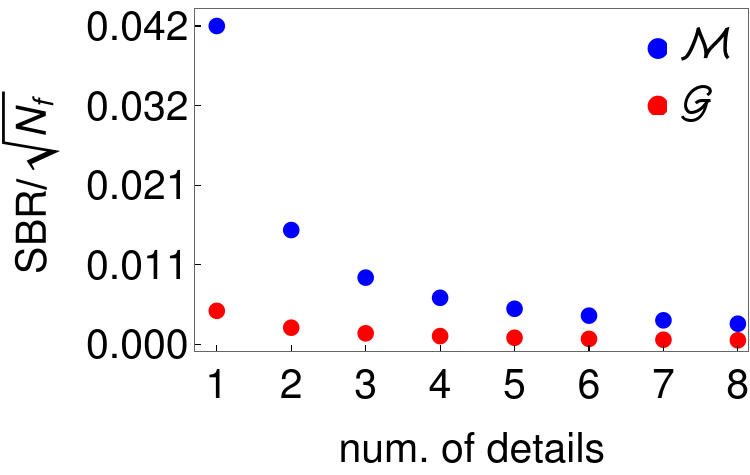}
	\caption{Comparison of the two SBR in the deeply out-of-focus regime for the correlation functions $\Gamma(\br_a,0)$ obtained in setups $\mathcal{G}$ (red dots) and $\mathcal{M}$ (blue dots), by varying the number of details (Gaussian slits) of an object with transmittance profile \eqref{eq:object-1}.
	}
	\label{fig:corr-num_slit-2}
\end{figure}

\begin{figure}
    \includegraphics[width=0.4\textwidth]{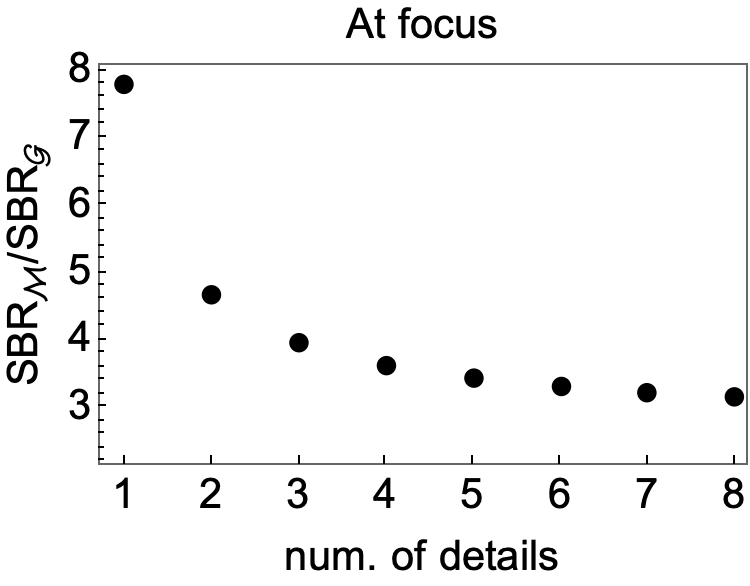}
    \hspace{0.03\textwidth}
	\includegraphics[width=0.4\textwidth]{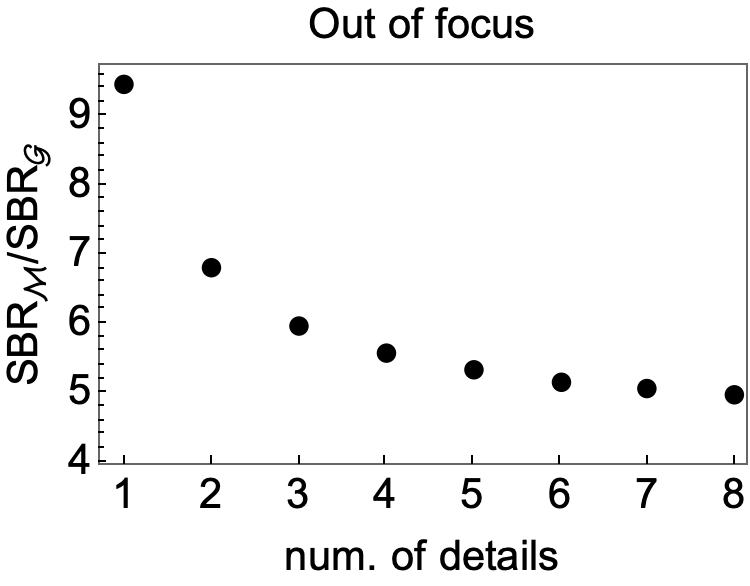}
	\caption{Ratio between the SBR of the correlation functions $\Gamma(\br_a,0)$ obtained in setup $\mathcal{M}$ and $\mathcal{G}$, by varying the number of details (Gaussian slits) of an object with transmittance profile \eqref{eq:object-1}. The left panel refers to the case at focus, while the right panel to the considered out-of-focus case. }
	\label{fig:corr-num_slit-1}
\end{figure}

We also report simulations, at focus and out of focus, with the objects defined in Eq.~\eqref{eq:object-1} and characterized by a varying number of slits, choosing the same  parameters as in Section~\ref{subsec:Choice-of-parameters}. The apertures defining the resolution at focus have been set to be equal in the two setups, together with all the other parameters. Simulations at focus, reported in Figure~\ref{fig:corr-num_slit}, were performed on an object with details close to the diffraction limit. In the out-of-focus regime, whose results are reported in Figure~\ref{fig:corr-num_slit-2}, we placed the object at a distance from the focused plane equal to 30 times the depth of focus in both setups, with the object parameters adjusted in such a way to have correlation images encoded in $\Gamma(\br_a,\br_b=0)$ resolved at 99\% visibility. Finally, Figure~\ref{fig:corr-num_slit-1} shows the ratios between the SBR obtained in $\mathcal{M}$ and $\mathcal{G}$, in the cases at focus and out of focus.

\section{Discussion}\label{sec:discussion}

We have compared the performances, in terms of signal-to-noise and signal-to-background ratios, of two setups for correlation plenoptic imaging. Both schemes are characterized by the 3D imaging capability typical of plenoptic devices. However, in setup $\mathcal{G}$, images of the scene can be retrieved only by measuring intensity correlations, while setup $\mathcal{M}$ can even work as an ordinary microscope, if the object is placed in the main lens focal plane.

The analysis we performed shows a solid advantage of $\mathcal{M}$ in terms of SNR and SBR, in both cases of the correlation function, representing a set of viewpoints on the scene, and the refocused image, in which all perspectives are properly merged. The case of the SBR computed in setup $\mathcal{M}$ for the refocused image of an $N$-slit object is particularly striking, since it shows that the capability to discriminate the signal (i.e., the slit peaks) from the background increases with the number of slits, as opposed to the decreasing behavior observed in the case of $\mathcal{G}$. The SNR behaves as expected in all cases, with an overall decreasing trend with increasing number of object details, showing an advantage of $\mathcal{M}$ over $\mathcal{G}$ by a factor that is small in the focused case, which is only moderately interesting, but can range from 3 to 9 in the much more relevant out-of-focus case, where it tends to coincide with SBR.

The differences found in terms of SNR and SBR are noticeable especially in the perspective of reducing the acquisition times. Actually, an increase by factor $F$ in the SNR at a fixed $N_f$ produces a quadratic decrease by a factor $F^2$ in the required number of frames (hence, in the acquisition time) to reach a specific SNR value, targeted in an experiment or an application. Though the experimental measurements reported in Ref.~\cite{cpi_exp} and Ref.~\cite{cpm_exp} are not matched in terms of imaging parameters, it is evident that the quality of images in the latter case, based on a setup of the type $\mathcal{M}$, underwent an outstanding improvement compared to the former, in which setup $\mathcal{G}$ was used, notwithstanding the number of frames, smaller of at least one order of magnitude in Ref.~\cite{cpm_exp}.

The characterization of fluctuations that affect the measurement of correlation functions can provide an interesting basis to develop new measurement protocols, in which noisy contributions, that do not bring relevant information on images, can be erased \textit{a priori} in a deterministic way. A similar task was performed in the case of ghost imaging \cite{ferri_dgi} and \cite{,fluc_img1,fluc_img2}. Future research will be devoted to the generalization of the procedure to different correlation plenoptic imaging architectures.

\section{Methods}\label{sec:methods}

	The deterministic propagation from the source to the detectors along the two optical paths $a$ and $b$ depends on the transmission functions of the object and the lenses. We will consider the field modelled by a scalar function $V$, related with the intensity by $I=|V|^2$, assuming that polarization degrees of freedom are relevant in the considered systems. In free space, a monochromatic field with frequency $\omega$ and wavenumber $k=\omega/c$, evaluated on a plane at a general longitudinal position $z$, is related with the same field at $z_{0}<z$ by the paraxial transfer function: 
	\begin{equation}
		V(\boldsymbol{\rho};z)=\frac{-\mathrm{i}k}{2\pi(z-z_{0})}\int \mathrm{d}^{2}\boldsymbol{\rho}^{\prime}V(\boldsymbol{\rho}^{\prime};z_{0})\mathrm{e}^{\mathrm{i}k\left[\frac{(\boldsymbol{\rho}-\boldsymbol{\rho}^{\prime})^{2}}{2(z-z_{0})}+(z-z_{0})\right]}.\label{free}
	\end{equation}
	We shall treat radiation emission by the chaotic source as an approximately Gaussian random process, stationary and ergodic. In particular, the field $V_{S}(\boldsymbol{\rho}_{s})$ at a point $\boldsymbol{\rho}_{s}$ on the source will be characterized by a Gaussian-Schell equal-time
	correlator \citep{mandel} 
	\begin{equation}
		W_{S}(\boldsymbol{\rho}_{s},\boldsymbol{\rho}_{s}^{\prime})=\left\langle V_{S}(\boldsymbol{\rho}_{s})V_{S}^{*}(\boldsymbol{\rho}_{s}^{\prime})\right\rangle =I_{s}\mathrm{e}^{-\frac{\boldsymbol{\rho}_{s}^{2}}{4\sigma_{i}^{2}}-\frac{\boldsymbol{\rho}_{s}^{\prime2}}{4\sigma_{i}^{2}}-\frac{(\boldsymbol{\rho}_{s}-\boldsymbol{\rho}_{s}^{\prime})^{2}}{2\sigma_{g}^{2}}},\label{WS}
	\end{equation}
	where $I_{s}$ is a constant in $\mathcal{G}$, and is the object intensity profile in $\mathcal{M}$; $\sigma_g$ is the transverse coherence length. We will consider sources characterized by negligible transverse coherence, accordingly replacing the mutual coherence function $\exp(-\boldsymbol{\rho}^{2}/2\sigma_{g}^{2})$ in the propagation integrals
	with $2\pi\sigma_{g}^{2}\delta^{(2)}(\boldsymbol{\rho})$. In this hypothesis, Eqs.~\eqref{free}-\eqref{WS}
	fully define the correlation function in Eq.~\eqref{Gamma1}

The refocused image can be expressed as
\begin{equation}
	\Sigma_{\mathrm{ref}}(\boldsymbol{\rho}_{a})=\left\langle \sigma_{(\alpha,\beta)}\left(\boldsymbol{\rho}_{a}\right)\right\rangle ,\label{Sigmaref}
\end{equation}
in terms of the physical observable
\begin{equation}
	\sigma_{(\alpha,\beta)}(\boldsymbol{\rho}_{a})=\int \mathrm{d}^{2}\boldsymbol{\rho}_{b}\Delta I_{A}\left(\alpha\boldsymbol{\rho}_{a}+\beta\boldsymbol{\rho}_{b}\right)\Delta I_{B}\left(\boldsymbol{\rho}_{b}\right).\label{sigmaref}
\end{equation}
We obtain information on the noise affecting the refocused images from the statistical fluctuations of the observable \eqref{sigmaref} around its average $\Sigma_{\mathrm{ref}}(\boldsymbol{\rho}_{a})$,
namely 
\begin{equation}\label{eq:F}
	\mathcal{F}\left(\boldsymbol{\rho}_{a}\right)= \left\langle \sigma_{(\alpha,\beta)}(\boldsymbol{\rho}_{a})^{2}\right\rangle -\left\langle \sigma_{(\alpha,\beta)}(\boldsymbol{\rho}_{a})\right\rangle ^{2} = \int \mathrm{d}^{2}\boldsymbol{\rho}_{b1} \int\mathrm{d}^{2}\boldsymbol{\rho}_{b2}\Phi(\boldsymbol{\rho}_{a},\boldsymbol{\rho}_{b1},\boldsymbol{\rho}_{b2}),
\end{equation}
with $\Phi$ a positive function related with the local fluctuations
of intensity correlations [compare with the definition \eqref{sigmaref}]. Being up to forth-order
in the intensity correlations, $\Phi$ contains up to eight-order
correlation terms between electric fields.

To compute the quantities \eqref{Sigmaref}-\eqref{eq:F}, it is necessary to evaluate
up to eight-point field correlators. By using the Gaussian
approximation, we will assume that Isserlis-Wick's theorem is valid
for the correlators that involve an equal number of $V$'s and $V^{*}$'s,
namely 
\begin{equation}
	\left\langle \prod_{j=1}^{n}V_{S}(\boldsymbol{\rho}_{j})V_{S}^{*}(\boldsymbol{\rho}_{j}')\right\rangle =\sum_{\mathrm{P}}\prod_{j=1}^{n}\left\langle V_{S}(\boldsymbol{\rho}_{j})V_{S}^{*}(\mathrm{P}\boldsymbol{\rho}_{j}')\right\rangle ,\label{Wick}
\end{equation}
with $\mathrm{P}$ a permutation of the primed indexes, while all
other correlators, including $\left\langle V\right\rangle $ itself,
vanish.

Since field propagation from the source to the detectors is purely deterministic, the eight-point expectation values in \eqref{eq:F},
involving the fields at the detectors $A$ and $B$, are determined
by the eight-point correlators of the field at the source. In both
setups, the most relevant term in $\mathcal{F}(\boldsymbol{\rho}_{a})$
is determined by the only contribution to $\Phi(\boldsymbol{\rho}_{a},\boldsymbol{\rho}_{b1},\boldsymbol{\rho}_{b2})$
in \eqref{eq:F} that is factorized with respect to the fields at
the two detectors: 
\begin{equation}\label{F0}
	\mathcal{F}_{0}(\boldsymbol{\rho}_{a}):=\int \mathrm{d}^{2}\boldsymbol{\rho}_{b1} \int\mathrm{d}^{2}\boldsymbol{\rho}_{b2} \Gamma_{AA}\left(\alpha\boldsymbol{\rho}_{a}+\beta\boldsymbol{\rho}_{b1},\alpha\boldsymbol{\rho}_{a}+\beta\boldsymbol{\rho}_{b2}\right) \Gamma_{BB}\left(\boldsymbol{\rho}_{b1},\boldsymbol{\rho}_{b2}\right).
\end{equation}
The self-correlations of intensity fluctuations are defined, in analogy
to \eqref{Gamma1}, as 
\begin{equation}
\Gamma_{DD}(\boldsymbol{\rho}_{d1},\boldsymbol{\rho}_{d2})=\left\langle \Delta I_{D}(\boldsymbol{\rho}_{d1})\Delta I_{D}(\boldsymbol{\rho}_{d2})\right\rangle 
\end{equation}
with $D=A,B$. Actually, the integrand of \eqref{F0} is concentrated
around $\boldsymbol{\rho}_{b1}-\boldsymbol{\rho}_{b2}=0$ (see also \cite{cpi_snr}), but shows no concentration on the $\boldsymbol{\rho}_{b1}$ plane. Instead, the other terms that characterize $\Phi(\boldsymbol{\rho}_{a},\boldsymbol{\rho}_{b1},\boldsymbol{\rho}_{b2})$
show concentration in \textit{both} variables $\boldsymbol{\rho}_{b1}$ and
$\boldsymbol{\rho}_{b2}$. This property is reflected in the fact
that $\mathcal{F}-\mathcal{F}_{0}$ is typically suppressed with respect
to $\mathcal{F}_{0}$ like the ratio of the intensity of light contained
a coherence area on $D_{B}$ and the total intensity transmitted
to $D_{B}$. This feature is demonstrated in Ref.~\cite{cpi_snr} for $\mathcal{G}$, and in the Appendix for $\mathcal{M}$. Hence, throughout the paper, we have approximated the variance of the measured correlations by the term $\mathcal{F}_{0}$.
 
\bmhead{Data availability statement}

The datasets generated during and/or analysed during the current study are either included in this published article or available from the corresponding author on reasonable request.

\bmhead{Acknowledgments}

GS is supported by QuantEra 2/2020 grant; GM, FP and MD acknowledge the support by Istituto Nazionale di Fisica Nucleare (INFN) projects PICS4ME and TOPMICRO, and project Qu3D, supported by the Italian Istituto Nazionale di Fisica Nucleare, the Swiss National Science Foundation (grant 20QT21$\_$187716 ``Quantum 3D Imaging at high speed and high resolution''), the Greek General Secretariat for Research and Technology, the Czech Ministry of Education, Youth and Sports, under the QuantERA programme, which has received funding from the European Union's Horizon 2020 research and innovation programme. GS thanks M. Pawłowski and R. Demkowicz-Dobrza\'nski for the support which led to the realization of the present work.

\begin{appendices}

\section[\appendixname~\thesection]{Exact expressions for the refocusing function}
		The refocusing function $\Sigma_{\text{ref}}$ for $\mathcal{M}$ reads
		\begin{align}
			\Sigma_{\text{ref}}\left(\br_{a}\right)= & \int \mathrm{d}^2 \br_{b}\Gamma\left(\alpha\br_{a}+\beta\br_{b},\br_{b}\right)\nonumber \\
			= & \left(2\pi\sigma_{g}^{2}I_{s}M\right)^{2}  \\ & \times \int\mathrm{d}^2 \br_{o1} \int\mathrm{d}^2 \br_{o2} \int\mathrm{d}^2 \br_{s1} \int\mathrm{d}^2 \br_{s2} \lvert A\left(\br_{s1}\right)\rvert^{2}\lvert A\left(\br_{s2}\right)\rvert^{2} \nonumber \\
			& \times P_{o}\left(\br_{o1}\right)P_{o}^{*}\left(\br_{o2}\right) \mathcal{P}\left(k\left[\frac{M\beta}{f_T}\left(\br_{o2}-\br_{o1}\right)+\frac{1}{z_a}\left(\br_{s2}-\br_{s1}\right)\right]\right)\nonumber \\
			& \times e^{-\frac{ik}{2}\left(\frac{1}{z_a}-\frac{1}{f_O}\right)\left(\br_{o2}^{2}-\br_{o1}^{2}\right)+\frac{ik\alpha}{f_T}\left(\br_{o2}-\br_{o1}\right)\cdot\br_{a}+\frac{ik}{z_a}\left(\br_{s2}\cdot\br_{o2}-\br_{s1}\cdot\br_{o1}\right)}.\label{eq:Sigma_ref}
		\end{align}
		where $P_O$ is the objective lens pupil function, $\mathcal{P}\left(\kappa\right)=\int d\rho\lvert P_{o}\left(\rho\right)\rvert^{2}e^{-i\kappa\cdot\rho}$ the Fourier transform of the objective intensity transmittance, and $M=f_T/f_O$ the magnification of focused objects. 
		
		The dominant term $\mathcal{F}_{0}$ of the variance $\mathcal{F}$ reads
		\begin{equation}
			\mathcal{F}_{0}\left(\br_{a}\right) = \int \mathrm{d}\br_{b1} \int\mathrm{d}\br_{b2}\Gamma_{AA}\left(\alpha\br_{a}+\beta\br_{b1},\alpha\br_{a}+\beta\br_{b2}\right)\Gamma_{BB}\left(\br_{b1},\br_{b2}\right)
		\end{equation}
		with $\Gamma_{AA}$ ($\Gamma_{BB}$) the intensity self-correlation between two points on the detector $D_a$ ($D_b$). Thus,
		\begin{align}
			\mathcal{F}_{0}\left(\br_{a}\right) = & \left(2\pi\sigma_{g}^{2}I_{s}M\right)^{4} \nonumber \\ & \times \int \mathrm{d}^2\br_{o4} \int \mathrm{d}^2\br_{o3} \mathcal{A}\left(\frac{k(\br_{o4}-\br_{o3})}{z_a} \right) \\
			& \times \int \mathrm{d}^2\br_{o2}\int \mathrm{d}^2\br_{o1}  \mathcal{A}^{*}\left(\frac{k(\br_{o2}-\br_{o1})}{z_a} \right)\nonumber \\
			& \times P_{o}^{*}\left(\br_{o3}\right)P_{o}\left(\br_{o4}\right)P_{o}^{*}\left(\br_{o2}\right)P_{o}\left(\br_{o1}\right) 
			\\ & \times\int \mathrm{d}^2\br_{s2} \int\mathrm{d}^2\br_{s1} \bigg\lvert A\left(\br_{s1}\right)\bigg\rvert^{2}\bigg\lvert A\left(\br_{s2}\right)\bigg\rvert^{2}\nonumber \\
			& \times \mathcal{P}\left(k\left[\frac{\beta M}{f_T}\left(\br_{o1}-\br_{o3}\right)+\frac{\br_{s1}-\br_{s2}}{z_a}\right]\right)
			\\ & \times \mathcal{P}\left(k\left[\frac{\beta M}{f_T}\left(\br_{o4}-\br_{o2}\right)+\frac{\br_{s1}-\br_{s2}}{z_a}\right]\right)\nonumber \\
			& \times e^{\frac{ik}{2}\left(\frac{1}{z_a}-\frac{1}{f_O}\right)\left(\br_{o4}^{2}-\br_{o3}^{2}-\br_{o2}^{2}+\br_{o1}^{2}\right)-\frac{ik\alpha}{f_T}\left(\br_{o4}-\br_{o3}-\br_{o2}+\br_{o1}\right)\cdot\br_{a}} , \label{eq:F_Sigma}
		\end{align}
		where $\mathcal{A}\left(\kappa\right)=\int d\rho\lvert A\left(\rho\right)\rvert^{2}e^{-i\kappa\cdot\rho}$ is the Fourier transform of the object intensity transmittance. The corresponding expressions for $\mathcal{G}$ are
		reported in Ref.~\cite{cpi_snr}.

\end{appendices}

\end{document}